# Towards an Intelligent Data Delivery Service

*Wen Guan*[1,*], *Tadashi Maeno*[2], Gancho Dimitrov[3], Brian Paul Bockelman[4], Torre Wenaus[2], Vakhtang Tsulaia[5], and *Nicolo Magini*[6]

[1]University of Wisconsin-Madison, Madison, USA
[2]Brookhaven National Laboratory, Upton, USA
[3]CERN, Meyrin, Switzerland
[4]University of Nebraska Lincoln, Nebraska, USA
[5]Lawrence Berkeley National Laboratory, Berkeley, USA
[6]Iowa State University, IA, USA

**Abstract.** The ATLAS Event Streaming Service (ESS) at the LHC is an approach to preprocess and deliver data for Event Service (ES) that has implemented a fine-grained approach for ATLAS event processing. The ESS allows one to asynchronously deliver only the input events required by ES processing, with the aim to decrease data traffic over WAN and improve overall data processing throughput. A prototype of ESS was developed to deliver streaming events to fine-grained ES jobs. Based on it, an intelligent Data Delivery Service (iDDS) is under development to decouple the "cold format" and the processing format of the data, which also opens the opportunity to include the production systems of other HEP experiments. Here we will at first present the ESS model view and its motivations for iDDS system. Then we will also present the iDDS schema, architecture and the applications of iDDS.

## 1 Introduction

The ATLAS [1] experiment at the LHC [2] [3] has accumulated more than 480 Petabytes of data processed in an internationally distributed Grid infrastructure with around 170 computing centers in more than 40 countries, which is capable of providing about 6M CPU-hours/day. Although this scale of computing facility is huge, ATLAS computing is still resource constrained. Furthermore, the data size and complexity is growing quickly with the operation of the experiment. The HL-LHC is scheduled to begin operation in 2026. When it happens, the produced data will increase significantly. By 2038, a factor of 20 more data than at present will be produced. With this ever-growing volume of data in the near future, the current LHC distributed computing model will lack resources [4]. To overcome this challenge, research and development is performed in different directions. One of the directions is to increase the efficiency of data usage with less expensive replicas. The ATLAS Event Streaming Service (ESS) [5] proposes a way to deliver fine-grained input data over the Wide Area Network (WAN), without creating expensive replicas of the input files. Based on ESS, a new service iDDS is proposed to intelligently transform and deliver the needed data to the processing workflow in a fine-grained approach. It will not only reduce the need for replicas, but also enable benefits such as decreasing the time period of caching transient data, transforming expensive replicas to cheaper format data at remote sites and only cache cheaper new data for processing. To avoid removing the transient data before it is processed, iDDS also needs to orchestrate the WorkFlow Management System (WFMS) and the Distributed Data Management (DDM) systems to trigger them to process the data as soon as possible. In addition, iDDS will have intelligent algorithms to adjust the lifetime of cache, the format transformation and the delivery destination. The iDDS will increase the efficiency of data usage, reduce storage usage for processing and speed up the processing workflow. The iDDS is currently under development. The first application is from ATLAS data carousel project, which aims to "increase the usage of less expensive storage" [6] and "orchestrate data processing between workload management, data management, and storage services with the bulk data resident on offline storage" [6]. An improved data carousel workflow to deliver fine-grained data from tape is already implemented and will be integrated into the ATLAS workflow management of PanDA and Prodsys2.

---
[*] email: wen.guan@cern.ch

## 2 Event Streaming Service (ESS)

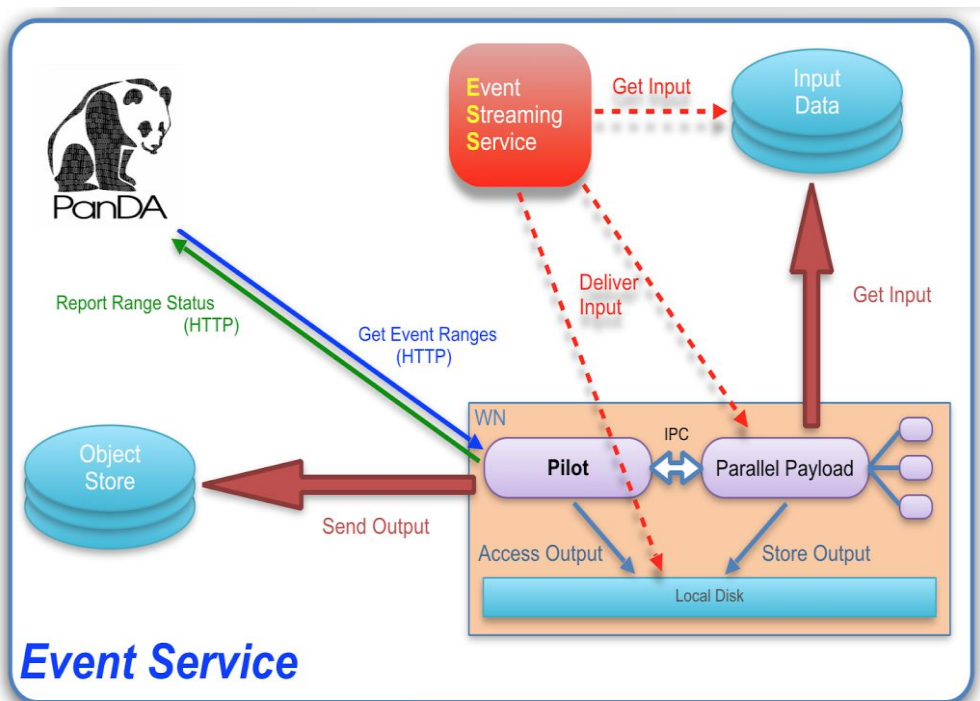

**Fig. 1.** A schematic view of the ATLAS Event Service and Event Streaming Service [3].

The ATLAS Event Service [7] [8], as shown in Figure 1, is a fine-grained approach to make maximal usage of opportunistic resources such as High Performance Computing (HPC) hole-filling, spot market commercial clouds, volunteer computing and shared grid resources. In the current Event Service implementation, processes running on the worker nodes require replicas of the full input files even though they only process a part of these events. This will become a limitation in the future, especially when requiring larger input datasets. To avoid creating these expensive replicas of the input files, ESS proposes pre-transformation of the input files at remote sites and only deliver required events to the application over Wide Area Network. This will reduce the replica usage and improve the efficiency by hiding WAN latency.

An ESS prototype has been designed to pre-fetch fine-grained events at the storage side and deliver the events to the workers just in time for processing. ESS introduces a way to reduce replica storage usage, and to improve the efficiency by hiding network latency.

## 3 Towards intelligent Data Delivery Service (iDDS)

Beyond ESS, several new workflows have recently been proposed. Below are some examples:

(1) For data carousel, instead of waiting to release jobs until all files of a dataset have been staged-in, we can process the file that is already staged-in and remove it after it's processed. In this fine-grained way, we can speed up the file processing and reduce the stage-in pool usage.

(2) For some analysis format data, such as DAOD, in the current computing model they are centrally produced and stored for a long time. Some of the produced data may never be used and some of them are used just a few times in a short period and are never touched after that. It occupies a lot of storage space. If we can produce these analysis format data on demand with adjustable lifetime based on the data usage frequency, we may be able to reduce the storage used by these data.

(3) For the HL-LHC, we will produce a lot more data and it will be difficult for users to download all data to local storage. However, some analysis methods such as machine learning require to load all events and then evaluate them in many iterations. As a result, a preparation step to skim/slim data and only download required event information to local storage is needed. If we can standardize this step and make it reusable, we will not only make the life of physicists easier, but may also reduce the CPU time used today to skim/slim the data many times for every analysis.

(4) Some existing services developed with accumulated requirements are housed in inappropriate components, for example, dynamic data placement service is mixed in the job management system PanDA and the data management system Rucio, which increases the difficulty to maintain them and to optimize the workflow.

Based on ESS, a new service iDDS is proposed to intelligently transform and deliver the needed data to a processing workflow in a high granularity. Here are the main functions of iDDS:

(1) Transformation on demand: Transform expensive data on demand to the format needed for processing on a remote site and only deliver the needed data to the following processing steps. At first, transformation on demand will avoid producing unused data. Secondly the storage-side transformation will minimize the network load. Thirdly, instead of delivering expensive complete replicas, only cheaper transformed data will be delivered and cached, which will reduce local replicas or cache usage. Last but not least, we can apply data locality knowledge and intelligence in the caching process to promote the cache reuse.

(2) Fine-grained delivery: Coordinate with the following processing steps to process data and to remove data in a fine-grained way, without waiting for all data to be cached. It will reduce the replica usage or cache usage and speed up the processing workflow.

(3) Orchestration: Orchestration between WFMS and DDM for optimal usage of limited resources for workflows that intersects the boundaries of data management and workflow management.

(4) Intelligent: To develop intelligent algorithms as a brain to improve the scheduling in iDDS, which will apply data locality knowledge and processing requests to trigger on-demand transformations, fine-grained delivery and cache management to optimize the processing workflow and promote the cache reuse.

## 4 The Design of Intelligent Data Delivery Service

The iDDS is designed as a standalone experiment agnostic service. It consists of a general Restful service to receive requests from WFMS and several running agents in a daemon mode to process the requests. The design schema is described in Figure 2.

In this model, the Restful service is used to register and query requests. It also provides a catalog service for users to retrieve required collections or contents. In the daemon mode, an agent Transporter works to find the input replicas from DDMs and another agent Transformer works to transform the expensive input replicas to desired format. When the output data is available, the agent Conductor works in a fine-grained approach to notify consumers to process the new transformed data.

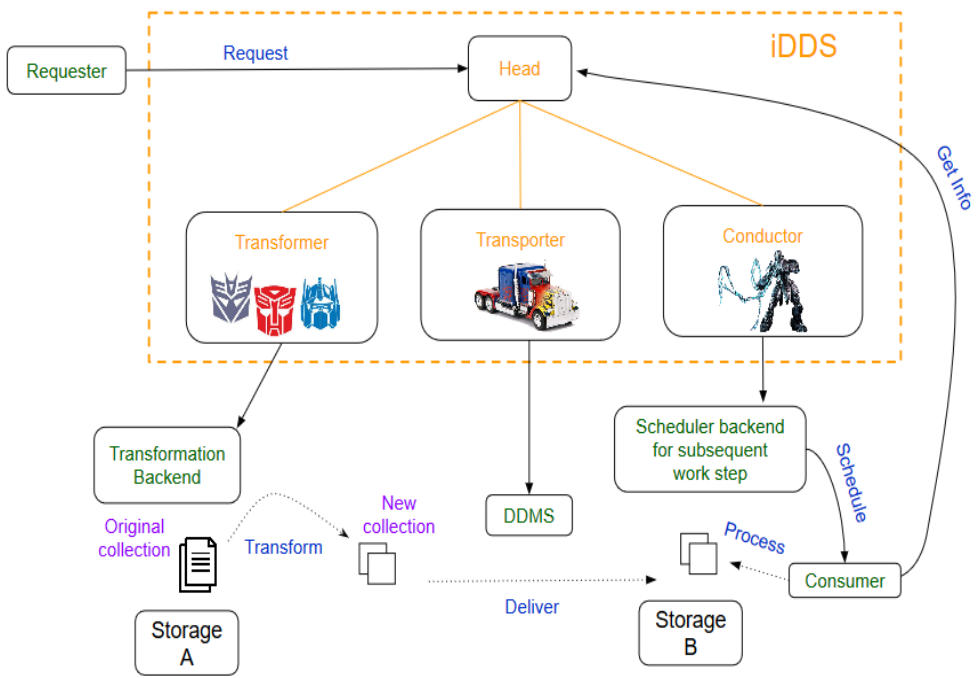

**Fig. 2.** A schematic view of the Intelligent Data Delivery Service.

### 4.1 The Architecture of iDDS

The iDDS is designed with abstract layers to hide the complexity of different logics and every layer concentrates on one type of operations, as shown in Figure 3. It simplifies the logic of every layer and smooths the development and maintenance. The iDDS is composed of ORM (Oject-Relational Mapping) layer, Core layer, API layer, Restful services, daemons and clients.

The iDDS plugin architecture is another experiment agnostic design to support new emerging workflows. In iDDS, a base plugin class is designed and external plugins can inherit from it to implement new workflow functions. With this structure, iDDS can support different transforms and different data management systems.

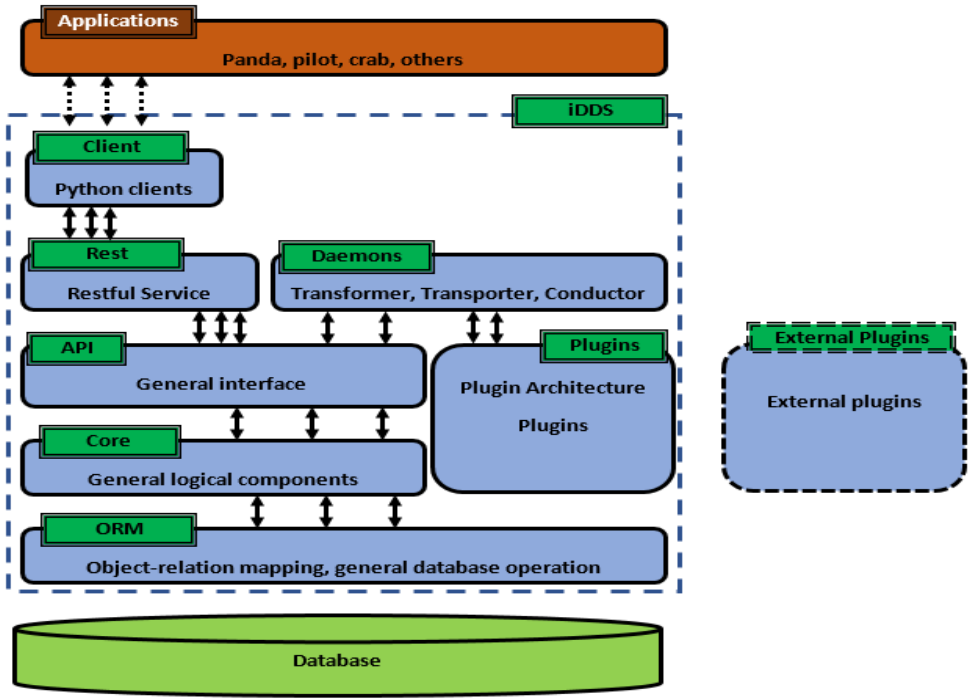

**Fig. 3.** A schematic view of the Architecture of Intelligent Data Delivery Service.

## 5 Outlook and Conclusions

We have implemented the main iDDS components and a group of plugins to support an ATLAS data carousel workflow. The integration tests are currently being performed, and more and more tasks will be injected to integrate iDDS with the ATLAS production system at the LHC. The next step will be to extend iDDS to support more workflows and to improve its intelligence.